\documentclass[12pt]{article}
\usepackage{amssymb}
\begin{document}

\title{\textbf{GENERAL RELATIVITY AND WEYL FRAMES }}
\author{C. ROMERO,\thanks{%
cromero@fisica.ufpb.br}{\ } J. B. FONSECA-NETO and M. L. PUCHEU}
\maketitle

\begin{abstract}
We present the general theory of relativity in the language of a
non-Riemannian geometry, namely, Weyl geometry. \ We show that the new
mathematical formalism may lead to different pictures of the same
gravitational phenomena, by making use of the concept of Weyl frames. We
show that, in this formalism, it is possible to construct a scalar-tensor
gravitational theory that is invariant with respect to the so-called Weyl
tranformations and reduces to general relativity in a particular frame, the
Riemann frame. In this approach the Weyl geometry plays a fundamental role
since it appears as the natural geometrical setting of the theory when
viewed in an arbitrary frame. Our starting point is to build an action that
is manifestly invariant with respect to Weyl transformations. When this
action is expressed in more familiar terms of Riemannian geometry we find
that the theory has some similarities with Brans-Dicke theory of gravity. We
illustrate this point with an example in which a known Brans-Dicke vacuum
solution may appear when reinterpreted in a particular Weyl frame.

\end{abstract}

keywords: {Weyl frames; conformal transformations; general relativity.}


address: {Departamento de F\'{\i}sica, Universidade Federal da Para\'{\i}ba, Jo\~{a}o Pessoa, PB 58059-970, Brazil}

\section{Introduction}

It is a very\ well known fact that the principle of general covariance has
played a major role in leading Einstein to the formulation of the theory of
general relativity.\cite{Norton} The idea underlying this principle is that
coordinate systems are merely mathematical constructions to\ conveniently
describe physical phenomena, and hence should not be an essential part of
the fundamental laws of physics. In a more precise mathematical language,
what is being required is that the equations of physics be expressed in
terms of intrinsic geometrical objects, such as scalars, tensors or spinors,
defined in the space-time manifold. This mathematical requirement is
sufficient to garantee the invariance of the form of the physical laws (or
covariance of the equations) under arbitrary coordinate transformations. In
field theories, one way of constructing covariant equations is to start with
an action in which the Lagrangian density is a scalar function of the
fields. In the case of general relativity, as we know, the covariance of the
Einstein equations is a direct consequence of the invariance of the
Einstein-Hilbert action.

A rather different kind of invariance that has been considered in some
branches of physics is invariance under conformal transformations. These
represent changes in the units of length and time that differ from point to
point in the space-time manifold. Conformal transformations were first
introduced in physics by H. Weyl in his attempt to formulate a unified
theory of gravitation and electromagnetism.\cite{Weyl} However, in order to
introduce new degrees of freedom to account for the electromagnetic field
Weyl had to assume that the space-time manifold is not Riemannian. This
extension consists of introducing an extra geometrical entity in the
space-time manifold, a 1-form field $\sigma $, in terms of which the
Riemannian compatibility condition between the metric $g$ and the connection 
$\Gamma $ is redefined. Then, a group of transformations, which envolves
both $g$ and $\sigma $,\ is defined by requiring that under these
transformations the new compatibility condition remain invariant. In a
certain sense, this new invariance group, which we shall call the group of
Weyl transformations, include the conformal transformations as subgroup.

It turns out that Einstein's theory of gravity in its original formulation
is not invariant neither under conformal transformations nor under Weyl
transformations. One reason for this is that the geometrical language of
Einstein`s theory is completely based on Riemannian geometry. Indeed, for a
long time general relativity has been inextricably associated with the
geometry of Riemann. Further developments, however, have led to the
discovery of different geometrical structures, which we might \ generically
call ``non-Riemannian" geometries, Weyl geometry being one of the first
examples. Many of these developments were closely related to attempts at
unifying gravity with electromagnetism.\cite{Goenner} While the new born
non-Riemannian geometries were invariably associated with new gravity
theories, one question that naturally arises is to what extent is Riemannian
geometry the only possible geometrical setting for the formulation of
general relativity. Our aim in this paper is to show that, surprisingly
enough, one can formulate general relativity using the language of a
non-Riemannian geometry, namely, the one known as Weyl integrable geometry.
In this formulation, general relativity appears as a theory in which the
gravitational field is described simultaneously by two geometrical fields:
the metric tensor and the Weyl scalar field, the latter being an essential
part of the geometry, manifesting its presence in almost all geometrical
phenomena, such as curvature, geodesic motion, and so on. As we shall see,
in this new geometrical setting general relativity exhibits a new kind of
invariance, namely, the invariance under Weyl transformations.

The paper is organized as follows. In Sec.~2, we present the basic
mathematical facts of Weyl geometry and the concept of Weyl frames. In
Sec.~3, we show how to recast general relativity in the language of Weyl
integrable geometry. In this formulation, we shall see that the theory is
manifestly invariant under the group of Weyl transformations. We proceed, in
Sec.~4, to obtain the field equations and interpret the new form of theory
with Brans-Dicke scalar-tensor gravity. We conclude with some final remarks
in Sec.~5.

\section{Weyl Geometry}

Broadly speaking, we can say that the geometry conceived by Weyl is a simple
generalization of Riemannian. Indeed, instead of postulating that the
covariant derivative of the metric tensor $g$ is zero, we assume the more
general condition\cite{Weyl}

\begin{equation}
\nabla _{\alpha }g_{\mu \nu }=\sigma _{\alpha }g_{\mu \nu }
\label{compatibility}
\end{equation}%
where $\sigma _{\alpha }$ \ denotes the components with respect to a local
coordinate basis $\left\{ \frac{\partial }{\partial x^{\alpha }}\right\} $
of a one-form field $\sigma \ $defined on $M$. This, in fact, represents a
generalization of the Riemannian condition of compatibility between the
connection $\nabla $ and $g,$ namely, the requirement that the length of a
vector remain unaltered by parallel transport \cite{Pauli}. If $\sigma $ $%
=d\phi ,$ where $\phi $ is a scalar field, then we have what is called\ an 
\textit{integrable Weyl geometry}. The set $(M,g,\phi )$ consisting of a
differentiable manifold $M$ endowed with a metric $g$ and a Weyl scalar
field $\phi $ $\ $is usually referred to as a \textit{Weyl frame}. It is
interesting to note that the Weyl condition (\ref{compatibility}) remains
unchanged\ when we go to another Weyl frame $(M,\overline{g},\overline{\phi }%
)$ by performing the following simultaneous transformations in $g$ and $\phi 
$:%
\begin{equation}
\overline{g}=e^{f}g  \label{conformal}
\end{equation}%
\begin{equation}
\overline{\phi }=\phi +f  \label{gauge}
\end{equation}%
where $f$ is a scalar function defined on $M$.\footnote{%
As one can easily check, that the Weyl compatibility condition between $%
\nabla $ and $g$ is equivalent to the Riemnanian compatibility between $%
\nabla $ and $\overline{g}=e^{-\phi }g.$}

Quite analogously to Riemannian geometry, the condition (\ref{compatibility}%
) is sufficient to determine\ the Weyl connetion\ $\nabla $\ in terms of the
metric $g$\ and the Weyl one-form field $\sigma .$ Indeed, a straightforward
calculation shows that one can express the components of the affine
connection with respect to an arbitrary vector basis completely in terms of
the components of $g$ and $\sigma $:%
\begin{equation}
\Gamma _{\mu \nu }^{\alpha }=\{_{\mu \nu }^{\alpha }\}-\frac{1}{2}g^{\alpha
\beta }[g_{\beta \mu }\sigma _{\nu }+g_{\beta \nu }\sigma _{\mu }-g_{\mu \nu
}\sigma _{\beta }]  \label{Weylconnection}
\end{equation}%
where $\{_{\mu \nu }^{\alpha }\}=$ $\frac{1}{2}g^{\alpha \beta }[g_{\beta
\mu ,\nu }+g_{\beta \nu ,\mu }-g_{\mu \nu ,\beta }]$ represents the
Christoffel symbols, i.e., the components of the Levi-Civita connection.%
\footnote{%
Throughout this paper our convention is that Greek indices take values from $%
0$ to $n-1$.}

A clear geometrical insight on the properties of Weyl parallel transport is
given by the following proposition: Let $M$ be a differentiable manifold
with an affine connection $\nabla $, a metric $g$ and a Weyl field of
one-forms $\sigma $. If $\nabla $ is compatible with $g$ in the Weyl sense,\
i.e. if (\ref{compatibility}) holds, then for any smooth curve $C=C(\lambda
) $ and any pair of two parallel vector fields $V$ and $U$ along $C,$ we
have 
\begin{equation}
\frac{d}{d\lambda }g(V,U)=\sigma (\frac{d}{d\lambda })g(V,U)
\label{covariantderivative}
\end{equation}%
where $\frac{d}{d\lambda }$ denotes the vector tangent to $C$ and $\sigma (%
\frac{d}{d\lambda })$ indicates the aplication of the 1-form $\sigma $ on $%
\frac{d}{d\lambda }$. (In a coordinate basis, putting $\frac{d}{d\lambda }=%
\frac{dx^{\alpha }}{d\lambda }\frac{\partial }{\partial x^{\alpha }},$ $%
V=V^{\mu }\frac{\partial }{\partial x^{\mu }},U=U^{\nu }\frac{\partial }{%
\partial x^{\nu }},\sigma =\sigma _{d}dx^{d},$ the above equation reads $%
\frac{d}{d\lambda }(g_{\alpha \mu }V^{\alpha }U^{\mu })=\sigma _{\nu }\frac{%
dx^{\nu }}{d\lambda }g_{\alpha \mu }V^{\alpha }U^{\mu }.$)

If we integrate the equation (\ref{covariantderivative})\ along the curve $C$%
, starting from a point $P_{0}=C(\lambda _{0}),$ then we obtain%
\begin{equation}
g(V(\lambda ),U(\lambda ))=g(V(\lambda _{0}),U(\lambda
_{0}))e^{\int_{\lambda _{0}}^{\lambda }\sigma (\frac{d}{d\rho })d\rho }
\label{integral}
\end{equation}%
Putting $U=V$ and denoting by $L(\lambda )$ the length of the vector $%
V(\lambda )$ at an arbitrary point\ $P=C(\lambda )$\ of the curve, then it
is easy to see that in a local coordinate system $\left\{ x^{\alpha
}\right\} $ the equation (\ref{covariantderivative}) reduces to 
\[
\frac{dL}{d\lambda }=\frac{\sigma _{\alpha }}{2}\frac{dx^{\alpha }}{d\lambda 
}L 
\]

Consider the set of all closed curves $C:[a,b]\in R\rightarrow M$, i.e, with 
$C(a)=C(b).$ Then, \ we have the equation 
\[
g(V(b),U(b))=g(V(a),U(a))e^{\int_{a}^{b}\sigma (\frac{d}{d\lambda })d\lambda
}. 
\]%
It follows from Stokes' theorem that if $\sigma $ is an exact form, that is,
\ if there exists a scalar function $\sigma $, such that $\sigma =d\phi $,
then

\[
\oint \sigma (\frac{d}{d\lambda })d\lambda =0 
\]%
for any loop. In other words, in this case the integral $e^{\int_{\lambda
_{0}}^{\lambda }\sigma (\frac{d}{d\rho })d\rho }$ does not depend on the
path.\ 

Let us conclude this section with a few historical comments on Weyl
gravitational theory. Weyl developed an entirely new geometrical framework
to formulate his theory, the main goal of which was to unify gravity and
electromagnetism. As is well known, although admirably ingenious, Weyl's
gravitational theory turned out to be unacceptable as a physical theory, as
was immediately realized by Einstein, who raised objections to the theory.%
\cite{Pauli,Pais}  Einstein's argument was that in a non-integrable\ Weyl
geometry the existence of sharp spectral lines in the presence of an
electromagnetic field would not be possible since atomic clocks would depend
on their past history.\cite{Pauli} However, it has been shown that a variant
of Weyl geometries, known as Weyl integrable geometry, does not suffer from
the drawback pointed out by Einstein. Indeed, it is the integral $%
I(a,b)=\int_{a}^{b}\sigma (\frac{d}{d\lambda })d\lambda $ that is
responsible for the difference between the readings of two identical atomic
clocks following different paths. Because in Weyl integrable geometry $%
I(a,b) $ is not path-dependent it has attracted the attention of many
cosmologists in recent years.\cite{Novello}

\section{General Relativity and a New Kind of Invariance}

We have seen in the previous section that the Weyl compatibility condition (%
\ref{compatibility}) is preserved\ when we go from a frame $(M,g,\phi )$ to
another frame $(M,\overline{g},\overline{\phi })$ through the
transformations (\ref{conformal}) and (\ref{gauge}). This has the
consequence that the components $\Gamma _{\mu \nu }^{\alpha }$ of the affine
connection are invariant under Weyl transformations, which, in turn, implies
the invariance of the affine geodesics. Now, as is well known, geodesics%
\textit{\ }plays a fundamental role in general relativity (GR) as well as in
most metric theories of gravity. Indeed, an elegant aspect of the
geometrization of the gravitational field lies in the geodesics postulate,
i.e., the statement that light rays\ and particles moving under the
influence of gravity alone follow space-time geodesics. Therefore a great
deal of information about the motion of particles in a given space-time is
promptly available once one knows its geodesics. The fact that geodesics are
invariant under (\ref{conformal}) and (\ref{gauge}) and that Riemannian
geometry is a particular case of Weyl geometry seems to suggest that it
should be possible to express general relativity in a more general
geometrical setting, namely, one in which the form of the field equations is
also invariant under Weyl transformations. In this section, we shall show
that this is indeed possible, and we shall proceed through the following
steps. First we shall assume that the space-time manifold which represents
the arena of physical phenomena may be described by a Weyl integrable
geometry, which means that now gravity will be described by two geometric
entities: a metric and a scalar field. The second step is to set up an
action $S$ invariant under Weyl transformations. We shall require that $S$
be chosen such that there exists a unique frame in which it reduces to the
Einstein-Hilbert action. The third step consists of extending Einstein's
geodesic postulate to arbitrary frames, such that in the Riemann frame it
should describe the motion of test particles and light exactly in the same
way as predicted by general relativity. Finally, the fourth step is to
define proper time in an arbitrary frame. This definition should be
invariant under Weyl transformations and coincide with the definition of
GR's proper time in the Riemann frame. It turns out then that the simplest
action that can be built under these conditions is%
\begin{equation}
S=\int d^{4}x\sqrt{-g}e^{-\phi }\left\{ R+2\Lambda e^{-\phi }+\kappa
e^{-\phi }L_{m}\right\} ,
\end{equation}%
where $R$ denotes the scalar curvature defined in terms of the Weyl
connection, $\Lambda $ is the cosmological constant, $L_{m}$ stands for the
Lagrangian of the matter fields and $\kappa $ is the Einstein`s constant. In 
$n$-dimensions we would have 
\begin{equation}
S_{n}=\int d^{n}x\sqrt{-g}e^{\left( 1-\frac{n}{2}\right) \phi }\left\{
R+2\Lambda e^{-\phi }+\kappa e^{-\phi }L_{m}\right\} .  \label{action1}
\end{equation}

In order to see that the above action is, in fact, invariant with respect
to\ Weyl transformations, we just need to recall that under (\ref{conformal}%
) and (\ref{gauge}) we have $\overline{g}^{\mu \nu }=e^{-f}g^{\mu \nu }$, $%
\sqrt{-\overline{g}}=e^{\frac{n}{2}f}\sqrt{-g}$, $\overline{R}_{\;\nu \alpha
\beta }^{\mu }=R_{\;\nu \alpha \beta }^{\mu },$ $\overline{R}_{\mu \nu
}=R_{\mu \nu },$ $\overline{R}=\overline{g}^{\alpha \beta }\overline{R}%
_{\alpha \beta }=e^{-f}g^{\alpha \beta }R_{\alpha \beta }=$ $e^{-f}R$. It
will be assumed that $L_{m}$ generally depends on $\phi ,$ $g_{\mu \nu }$
and the matter fields, its form being obtained from the special theory of
relativity through the prescription $\eta _{\mu \nu }\rightarrow e^{-\phi
}g_{\mu \nu }$ and $\partial _{\mu }\rightarrow \nabla _{\mu }$, where $%
\nabla _{\mu }$ denotes the covariant derivative with respect to the affine
connection. As it can be easily seen, these rules assure the invariance
under Weyl transformations of part of the action that is responsible for the
coupling of matter with the gravitational field, and, at the same time,
reduce to the principle of minimal coupling adopted in general relativity
when we set $\phi =0$, that is, in the Riemann frame.

We now turn our attention to the motion of test particles and light rays.
Here, our task is to extend GR's geodesic postulate in such a way that it is
invariant under Weyl transformations. The extension is straightforward and
may be stated as follows: if we represent parametrically a timelike curve as 
$x^{\mu }=x^{\mu }(\lambda )$, then this curve\ will correspond to the world
line of a particle free from all non-gravitational forces, passing through
the events $x^{\mu }(a)$ and $x^{\mu }(b)$, if and only if it extremizes the
functional 
\begin{equation}
\Delta \tau =\int_{a}^{b}e^{-\frac{\phi }{2}}\left( g_{\mu \nu }\frac{%
dx^{\mu }}{d\lambda }\frac{dx^{\nu }}{d\lambda }\right) ^{\frac{1}{2}%
}d\lambda ,  \label{propertime}
\end{equation}%
which is obtained from the special relativistic expression of proper time by
using the prescription $\eta _{\mu \nu }\rightarrow e^{-\phi }g_{\mu \nu }.$
Clearly, the right-hand side of this equation is invariant under Weyl
transformations and reduces to the known expression of the propertime in
general relativity in the Riemann frame. We take $\Delta \tau $, as given
above, as the extension to an arbitrary Weyl frame the clock hypothesis,
i.e., the assumption that $\Delta \tau $ measures the proper time measured
by a clock attached to the particle.\cite{Mainwaring}

It is not difficult to verify that the extremization condition of the
functional (\ref{propertime}) leads to the equations 
\[
\frac{d^{2}x^{\mu }}{d\lambda ^{2}}+\left( \left\{ _{\alpha \beta }^{\mu}\right\} -
\frac{1}{2}g^{\mu \nu }(g_{\alpha \nu }\phi _{,\beta }+
g_{\beta\nu }\phi _{,\alpha }-
g_{\alpha \beta }\phi \,_{,\nu }\right) \frac{dx^{\alpha }}{d \lambda }\frac{dx^{\beta }}{d\lambda }=0  
\]
where $\left\{ _{\alpha \beta }^{\mu }\right\} $ denotes the Christoffel
symbols calculated with $g_{\mu \nu }$. Let us recall that in the derivation
of the above equations the parameter $\lambda $ has choosen such that%
\begin{equation}
e^{-\phi }g_{\alpha \beta }\frac{dx^{\alpha }}{d\lambda }\frac{dx^{\beta }}{%
d\lambda }=K=const.  \label{constant}
\end{equation}%
along the curve, which, up to an affine transformation, permits the
identification of $\lambda $ with the proper time $\tau $. It turns out that
these equations are exactly those that yield the affine geodesics in a Weyl
integrable space-time as they can be rewritten as 
\begin{equation}
\frac{d^{2}x^{\mu }}{d\tau ^{2}}+\Gamma _{\alpha \beta }^{\mu }\frac{%
dx^{\alpha }}{d\tau }\frac{dx^{\beta }}{d\tau }=0,  \label{Weylgeodesics}
\end{equation}%
where $\Gamma _{\alpha \beta }^{\mu }=\left\{ _{\alpha \beta }^{\mu
}\right\} -\frac{1}{2}g^{\mu \nu }(g_{\alpha \nu }\phi _{,\beta }+g_{\beta
\nu }\phi _{,\alpha }-g_{\alpha \beta }\phi \,_{,\nu })$, according to (\ref%
{Weylconnection}), may be identified\ with the Weyl connection when $\sigma
_{\alpha }=$ $\phi _{,\alpha }$. Therefore, the extension of the geodesic
postulate by requiring that the functional (\ref{propertime}) be an extremum
is equivalent to postulating that the particle motion must follow affine
geodesics defined by the Weyl connection $\Gamma _{\alpha \beta }^{\mu }$.
It will be noted that, as a consequence of the Weyl compatibility condition (%
\ref{compatibility}) between the connection and the metric, (\ref{constant})
holds automatically along any affine geodesic determined by (\ref%
{Weylgeodesics}). Because both the connection components $\Gamma _{\alpha
\beta }^{\mu }$ and the proper time $\tau $ are invariant when we switch
from one Weyl frame to the other, the equations (\ref{Weylgeodesics}) are
manifestly covariant\ under Weyl transformations.\ 

As we know, the geodesic postulate not only makes a statement with respect
to the motion of particles, but also regulates the propagation of light rays
in space-time. Because the path of light rays are null curves, one cannot
use proper time as a parameter to describe them. In fact, light rays are
supposed to follow null affine geodesics, which cannot be defined in terms
of the functional (\ref{propertime}), but, instead, they must be
characterized by their behaviour with respect to parallel transport. We
shall extend this postulate by simply assuming that light rays follow Weyl
null affine geodesics.

It is well known that null geodesics are preserved under conformal
transformations, although one needs to reparametrize the curve in the new
gauge. In the case of Weyl transformations null geodesics are also invariant
with no need of reparametrization, since, again, the connection components $%
\Gamma _{\alpha \beta }^{\mu }$ do not change under (\ref{conformal}) and (%
\ref{gauge}), while the condition (\ref{constant}) is obvioulsy not altered.
As a consequence, the causal structure of space-time remains unchanged in
all Weyl frames. This seems to complete our program of formulating general
relativity in a geometrical setting that exhibits a new kind of invariance,
namely, that with respect to Weyl transformations.

\section{General Relativity as a Scalar-Tensor Theory}

In the new formalism it is interesting to rewrite the action (\ref{action1})
in Riemannian terms. This is done by expressing the Weyl scalar curvature $R$
in terms of the Riemannian scalar curvature $\widetilde{R}$ and the scalar
field $\phi $, which gives 
\[
R=\widetilde{R}-(n-1)\square \phi +\frac{(n-1)(n-2)}{4}g^{\mu \nu }\phi
_{,\mu }\phi _{,\nu } 
\]%
where $\square \phi $ denotes the Laplace-Beltrami.\ \ It is easily shown
that, by inserting $R$ as given above into Eq.~(\ref{action1}) and using
Stokes' theorem to neglect divergence terms in the integral, one obtains 
\begin{equation}
S_{n}=\int d^{n}x\sqrt{-g}e^{\left( 1-\frac{n}{2}\right) \phi }\left\{ 
\widetilde{R}+\omega g^{\mu \nu }\phi _{,\mu }\phi _{,\nu }+2\Lambda
e^{-\phi }+\kappa e^{-\phi }L_{m}\right\} ,  \label{action2}
\end{equation}%
where $\omega =\frac{(n-1)(2-n)}{4}$. For $n=4$ we have $\omega =-\frac{3}{2}
$ and the action becomes 
\begin{equation}
S=\int d^{4}x\sqrt{-g}e^{-\phi }\left\{ \widetilde{R}-\frac{3}{2}g^{\mu \nu
}\phi _{,\mu }\phi _{,\nu }+2\Lambda e^{-\phi }+\kappa e^{-\phi
}L_{m}\right\} .  \label{action3}
\end{equation}

At this point, it is convenient to change the \ scalar field variable $\phi $
by defining $\Phi =e^{-\phi }$. In terms of the new field $\Phi $ the action
(\ref{action3}) takes the form 
\begin{equation}
S=\int d^{4}x\sqrt{-g}\left\{ \Phi \widetilde{R}-\frac{3}{2\Phi }g^{\mu \nu
}\phi _{,\mu }\phi _{,\nu }+2\Lambda \Phi ^{2}+\kappa \Phi ^{2}L_{m}\right\}
.  \label{action4}
\end{equation}%
We then see that in the vacuum case and vanishing cosmological constant (\ref%
{action4}) is identical to the action of Brans-Dicke theory for $\omega =-%
\frac{3}{2}$. \ This fact has been pointed out by some authors in similar
contexts, although it will be noted that the analogy is not perfect since in
the Weyl frame the geodesics are not Riemannian.\cite{Dabrowski,Deruelle}

Finally, the field equations\ in an arbitrary Weyl frame are obtained by
taking variations of $S$ in (\ref{action3}) with respect to $g_{\mu \nu }$
and $\Phi $, these being considered\ as independent fields. This will give
us, respectively, 
\begin{equation}
\widetilde{G}_{\mu \nu }-(\phi _{,\mu ;\nu }-g_{\mu \nu }\square \phi )-%
\frac{1}{2}(\phi _{,\mu }\phi _{,\nu }+\frac{1}{2}g_{\mu \nu }\phi _{,\alpha
}\phi ^{,\alpha })=-e^{\phi }(\kappa T_{\mu \nu }-\Lambda g_{\mu \nu }) 
\label{g}
\end{equation}%
\begin{equation}
\widetilde{R}-3\square \phi +\frac{3}{2}\phi _{,\alpha }\phi ^{,\alpha
}=e^{\phi }(\kappa T-4\Lambda ),  \label{fi}
\end{equation}%
where $\widetilde{G}_{\mu \nu }$ denotes the Einstein tensor calculated with
the Riemannian connection, and $T=g^{\mu \nu }T_{\mu \nu }.$\ It should be
noted that (\ref{fi}) is just the trace of (\ref{g}), and so, the field
equations are not independent. This is consistent with the fact that we have
complete freedom in the choice of the Weyl frame. It also means that $\phi $
may be viewed as an arbitrary function and not a dynamical field.

\section{Final Remarks}

As we have seen, in this scenario the gravitational field is not associated
only with the metric tensor, but with the combination of both the metric $%
g_{\mu \nu }$ and the geometrical scalar field $\phi $. We can get some
insight in the amount of physical information carried by the scalar field by
investigating conformally-flat solutions of general relativity. Consider,
for instance, homogeneous and isotropic cosmological models. All these have
a conformally-flat geometry. It implies that there is a frame in which the
geometry of these models becomes that of flat Minkowski space-time. Thus, in
the Riemann frame the space-time manifold is endowed with a metric that
leads to Riemannian curvature, while in the Weyl frame space-time is flat.
In this case, we see that all information about the gravitational field is
encoded by the scalar field. Clearly, this leads, in distinct frames, to
different pictures of the same gravitational phenomena. For instance,
consider O`Hanlon-Tupper vacuum solution in Brans-Dicke theory with $\omega
=-\frac{3}{2}$.\cite{O'Hanlon} Surely, as far as the metric is concerned it
can be regarded as a solution of general relativity in a Weyl frame. In
fact, it is equivalent to Minkowski space-time in the Riemann frame,
although test particles follow affine geodesics, which do not coincide with
the metric geodesics of Minkowski space-time

An important conclusion to be drawn from what has been shown in this paper
is that general relativity can perfectly \textquotedblleft
survive\textquotedblright\ in a non-Riemannian environment. Moreover, as far
as physical observations are concerned, all Weyl frames, each one
determining a specific geometry, are completely equivalent. This conclusion
seems to give some support to the view conceived by H. Poincar\'{e} that the
geometry of space-time is perhaps a convention that can be freely chosen by
the theoretician.\cite{Henri} In particular, according to this view, general
relativity may be rewritten in terms an arbitrary conventional geometry.\cite%
{Tavakol}

Finally, it should be said that while at the classical level we have
complete equivalence of frames, in the quantum context this may not be true.
Indeed, as in the case of conformal transformations, quantization and Weyl
transformations may not\ always commute.\cite{Fujii}

\section*{Acknowledgments}

C.Romero and M. L.Pucheu would like to thank CNPq/CLAF for financial support.


\section{References}

\begin{thebibliography}{99}
\bibitem{Norton} J. D. Norton, \textit{Rep. Prog. Phys.} \textbf{56}, 791
(1993).

\bibitem{Weyl} H. Weyl,\textit{Sitzungesber Deutsch. Akad. Wiss. Berli} 465
(1918); H. Weyl, \textit{Space, Time, Matter} (Dover, New York, 1952).

\bibitem{Goenner} H. Goenner, \textit{Living Rev. Rel.} \textbf{7}, 2 (2004).

\bibitem{Pauli} A nice account of Weyl's ideas as well as the refutation of
his gravitational theory may be found in W. Pauli, \textit{Theory of
Relativity} (Dover, New York, 1981). See, also, L. O'Raiefeartaigh and N.
Straumann,\textit{Rev. Mod. Phys.} \textbf{72}, 1 (2000).

\bibitem{Pais} A. Pais, \textit{Subtle is the Lord} (Oxford University
Press, 1983)

\bibitem{Novello} M. Novello and H. Heintzmann, \textit{Phys. Lett. A} 
\textbf{98}, 10 (1983); K. A. Bronnikov, Yu. M. Konstantinov and V. N.
Melnikov, \textit{Grav. Cosmolog.} \textbf{1}, 60 (1995);  M. Novello,
L.A.R. Oliveira, J.M. Salim and E. Elbas,  \textit{Int. J. Mod. Phys. D} 
\textbf{1} 641 (1993);  J. M. Salim and S. L. Saut\'{u},  \textit{Class.
Quant. Grav} \textbf{13}, 353 (1996); H. P. de Oliveira, J. M. Salim and S.
L. Saut\'{u}, \textit{Class. Quant. Grav.} \textbf{14}, 2833 (1997);  V.
Melnikov, \textit{Classical Solutions in Multidimensional Cosmology} in 
\textit{Proceedings of the VIII Brazilian School of Cosmology and
Gravitation II}, ed. M. Novello (Editions Fronti\`{e}res, 1995) p.542; O.
Arias, R. Cardenas and I.Quiros, \textit{Nucl. Phys. B} \textbf{643}, 187
(2002); J. Miritzis, \textit{Class. Quantum .Grav.} \textbf{21}, 3043
(2004); J. Miritzis, \textit{J.Phys.: Conf. Ser.} \textbf{8},131 (2005); M.
Israelit, \textit{Found. Phys.} \textbf{35}, 1725 (2005); F. Dahia, G. A. T.
Gomez and C. Romero, \textit{J. Math.Phys.} \textbf{49}, 102501 (2008); J.
E. Madriz Aguilar and C. Romero, \textit{Found. Phys.} \textbf{39}, 1205
(2009).

\bibitem{Mainwaring} S. R. Mainwaring and G. E. Stedman, \textit{Phys. Rev. A%
} \textbf{47}, 3611 (1993).

\bibitem{Dabrowski} M. P. Dabrowski, T. Denkiewicz and D. Blaschke, \textit{%
Annalen Phys.} \textbf{16}, 237 (2007).

\bibitem{Deruelle}  N. Deruelle and M. Sasaki,  Conformal equivalence in
classical gravity: the example of  ``vailed" General Relativity, 
arXiv:gr-qc/1007.3563 (2010).

\bibitem{O'Hanlon} J. O`Hanlon and B. Tupper, \textit{Nuovo Cimento B} 
\textbf{7}, 305 (1972).

\bibitem{Henri}  H. Poincar\'{e}, \textit{Science and Hypothesis} (Dover,
New York, 1952).

\bibitem{Tavakol} I. W. Roxburgh and R. K. Tavakol, \textit{Found. Phys.} 
\textbf{8}, 229 (1978).
\end{thebibliography}
\end{document}